\documentclass[manuscript]{aastex}

\begin{document}
\pagenumbering{arabic}

\title{THE FORMATION AND EVOLUTION OF MASSIVE 
STAR CLUSTERS: HISTORICAL OVERVIEW}

\author{Sidney van den Bergh}
\affil{Dominion Astrophysical Observatory, Herzberg Institute of Astrophysics, National Research Council of Canada, 5071 West Saanich Road, Victoria, BC, V9E 2E7, Canada}
\email{sidney.vandenbergh@nrc.ca}

\begin{abstract}                  
   Some factors connecting the evolutionary histories of galaxies
with the characteristics of their cluster systems are reviewed. 
Unanswered questions include: How is one to understand the 
observation that some globular cluster systems have disk kinematics 
whereas others do not? Why do some galaxies have cluster systems 
with unimodal metallicity distributions, whereas others have 
bimodal metallicity distributions? What caused the average
ellipticity of individual clusters to differ from galaxy to galaxy?
 \end{abstract} 

\section{INTRODUCTION}

\subsection{Why there are star clusters and associations?}

   The existence of multiple stars and star clusters is, as
Larson (2003) has recently reminded us, due to the fact that 
the angular momentum in typical dense cores of giant 
interstellar molecular clouds (GMCs) is three orders of magnitude 
greater than can be contained within a single star. The magnetic 
fields in GMCs cannot dispose of all of this angular momentum. 
This is so because it has become both dynamically unimportant, and 
decoupled from the gas, during the later stages of the collapse of 
these GMC cores. The dense cores of such molecular clouds are 
therefore expected to fragment into clumps of massive stars. Such
concentrations of young stars were named ``associations'' by Ambartsumian 
(1954). Associations are now known to come in three flavors: (1)
OB-associations, which are comprised of massive young main sequence 
stars, (2) T-associations, which consist of young intermediate-mass 
stars that are still contracting to the main sequence, and (3) 
R-associations (van den Bergh 1966), which are made up of stars that
illuminate the dusty interstellar clouds in which they are still 
embedded. Since the space density of R-associations is much greater 
than that of OB-associations such R-associatiations are particularly 
suitable for the mapping of nearby Galactic spiral structure (van den 
Bergh 1968).

\subsection{Structure of associations.}
   Within low-density expanding (positive energy) associations one 
often also also finds denser gravitationally stable (negative energy) 
star clusters (e.g. Blaauw 1964). Some of these clusters, like the 
small (R = 0.3 pc) clustering near o Persei (that is located in the 
emission nebula IC 348) are so small that they clearly have not yet 
expanded out to their eventual equilibrium radii. Not all associations 
have the same structure. The 30 Doradus association (see Fig. 1 of 
Walborn, Ma\'{i}z-Apell\'{a}niz \& Barb\'{a} 2002) in the Large Magellanic 
Cloud is centered on the single massive compact young OB cluster 
R136 (which forms the core of the 30 Doradus association), whereas the
comparably rich association NGC 604 in M33 (see Fig. 2 of Miskey \& 
Bruhweiler 2003) contains many smaller subclusterings, rather than a
single massive core. NGC 206, which is the most luminous association 
in the Andromeda galaxy (M31), is a slightly older example of a large
association that does not have a single massive core. It is presently 
not known why some associations exhibit such subclustering whereas 
others do not.

\section{STAR CLUSTERS AND GALACTIC EVOLUTION}

\subsection{Ages of halo and disk clusters.}
   Hansen et al. (2002) have found that the faintest white dwarfs 
in the globular cluster M 4 are 2.5 mag fainter than are the faintest
field white dwarfs in the Solar neighborhood. The resulting cooling
ages are 12.7 $\pm$  0.7 Gyr for M 4 and 7.3 $\pm
$  1.5 Gyr for the Galactic
disk. The relatively young age of the nearby disk has recently
been confirmed by Sandage et al. (2003) who determined the age
of old field subgiants from their {\it Hipparcos} parallaxes. These
results show, beyond reasonable doubt, that the Galactic disk is 
$\sim$5 Gyr younger than the Galactic halo. As was first emphasized by 
Berman \& Suchkov (1991) the only reasonable way to obtain such a 
large age difference is by stopping star formation via heating of
the Galactic gas to temperatures in excess of that defining virial
equilibrium. Such heating would likely have resulted from the
strong stellar wind generated by the burst of star formation
that accompanied the formation of the Galactic bulge. The fact
that the oldest open clusters, such as NGC 6791 and NGC 188,
have metallicities that are 1 - 2 times Solar shows that
these objects must have been formed from gas that had already
been greatly enriched by star and supernova formation.

\subsection{Cluster formation and galactic star formation rates.}

   Larsen \& Richtler (1999, 2000) have recently shown that the 
fraction of the total light of galaxies that is locked up in star 
clusters differs widely from galaxy to galaxy. Star clusters are,
for example, almost absent from the very inactive Local Group
galaxy IC 1613 (Baade 1963). On the other hand 5.6\% of the total
V-band light of the nearby starburst galaxy NGC 1569 is emitted by 
stars in clusters. If massive clusters are mainly formed during
intense star bursts then : (1) The history of cluster formation 
in a galaxy cannot be used as a proxy for the overall history of 
star formation in that galaxy, and (2) galaxies that exhibit an
above average specific cluster frequency (Harris \& van den Bergh 
1981) probably had at least one major starburst in their history.
In view of the apparent correlation between starbursts and the
formation of massive clusters it would be particularly interesting 
to see if the specific cluster frequency is enhanced in IC 10, 
which is the nearest starburst galaxy.
   It seems likely (Elmegreen \& Efremov 1997, Tan \& McKee 2002)
that massive clusters of all ages preferentially formed (and 
survived) in high pressure environments. This is so because virialized 
clouds are more tightly bound at high pressures than they are in low 
pressure environments. Such high pressure might be due to either a 
high background virial density or to large-scale shocks that are 
generated during galaxy collisions. It would be interesting to know 
if the great burst of globular cluster formation, that took place $\sim$12 
Gyr ago, was also triggered by collisions, or if shocks associated 
with the reionization of the Universe (See Miralda-Escud\'{e} 2003 for 
a review) might have triggered the great early burst of globular 
cluster formation.

\subsection{Cluster formation and galaxy interactions.}
   Van den Bergh (1960) found that the spiral arms of some late-type 
interacting galaxies were characterized by particularly patchy 
(DDO type Sc*) spiral arms. Since such patches are due to the 
presence of numerous star clusters and associations, this indicated 
that strong tidal interactions and mergers can trigger bursts of 
cluster formation. The notion that interactions between galaxies might 
enhance the formation rate of massive star clusters (e.g. Schweitzer 
1987) has lately received strong support from observations with the 
{\it Hubble Space Telescope} [see Whitmore (2003) for a recent review]. The 
fact that the fraction of all star formation that takes place in 
massive clusters is apparently enhanced during bursts of star formation 
indicates that the history of cluster formation in a particular galaxy 
might be quite different from the overall history of star formation in 
that same galaxy. In particular it appears likely that the specific 
cluster formation frequency was enhanced during the early phase of 
galaxy formation that was dominated by mergers. Proto-planetary disks
surrounding young stars might have a high probability of being 
disrupted in cluster stars than in field stars. It is therefore quite
possible that the fraction of stars that form planets may be reduced 
in early generations of star formation. However, it will be difficult 
to test this prediction because the low metallicity of the first 
generation of stars might also inhibit planet formation. A
possible counterexample is provided by the old planet that 
Sigurdsson et al. (2003) have recently found in the globular
cluster M 4. Nevertheless the fact that the disk fraction in
very young clusters decreases with age (Lada \& Lada 2003) does
show that the cluster environment can destroy protoplanetary
disks on a time scale of a few million years.

\subsection{Formation of star clusters and cluster location.}
   Morgan (1958, 1959) devised a new one-dimensional system of galaxy 
classification, in which the principal classification parameter 
was central concentration of light in the galaxy image. In the
footnotes to his main tables Morgan drew attention to the fact that
many spiral galaxies have knots of very active star formation just 
outside their nuclei. The Arches (age 2 Myr) and Quintuplet (age 4 Myr) 
are examples of such recently formed super star clusters in own Milky 
Way system. It is presently not clear why the formation of massive 
star clusters is favored in a region where differential rotation and
tidal stresses would appear to make the formation (and subsequent
gravitational collapse) of the massive cores giant molecular clouds 
particularly challenging. Barth et al. (1995) find that very
luminous star clusters may also occur at somewhat greater nuclear
distances in the star forming rings that surround some active active
galactic nuclei.

\subsection{Luminosity functions of stars in young star clusters.}

   Larson (1999) has recently reviewed the observational evidence 
on the luminosity function of individual stars in young star 
clusters. He concludes that ``No clear evidence has been found for 
any systematic dependence of the IMF [initial mass function] on any
property of the system studied, and this has lead to the current 
widely held view that the IMF is universal, at least in the local 
universe.'' However, the Arches cluster near the Galactic center 
(Figer et al. 1999) may have a flatter mass spectrum than typical 
young clusters in the Solar neighborhood. So, perhaps, the mass 
spectrum of star formation in clusters is not entirely independent 
of environment after all. 

\subsection{The mass spectrum of cluster formation.}

   In his book {\it ``Sternhaufen''}, which is the first monograph 
entirely devoted to star clusters, ten Bruggencate (1927)
divides clusters into three classes: (1) globular clusters,
(2) open clusters, and (3) star swarms, a class that would,
in modern terminology, consist of both moving groups and
stellar associations. Ten Bruggencate emphasized the fact
that globular clusters have a spheroidal distribution that
is centered on the Galactic bulge, whereas open clusters
are mainly located in the flattened disk of the Milky Way.
Furthermore he pointed to the differences between the color-
magnitude diagrams of these two types of clusters, which
we now know to be due to systematic differences in age and
chemical composition. Perhaps surprisingly, ten Bruggencate
did not discuss the systematic difference between the 
integrated luminosities of typical globular and galactic 
clusters. Modern work broadly supports the division of Galactic
star clusters into (1) old, luminous, metal-poor globular
clusters and (2) young, less luminous, and metal-rich open
clusters. However, the more extensive data base that is
presently available does show intermediate objects such
as (a) metal-rich globulars in the Galactic bulge, (b) a few 
open clusters such as NGC 6791 and NGC 188 with ages of 5-10 Gyr, 
and (c) some outer halo globular clusters with an ages on only 
$\sim$10 Gyr. Surprisingly the age dichotomy between globulars and 
open clusters is even more pronounced in the Large Magellanic
Cloud (Da Costa 2002) than it is in the Galaxy. On the other 
hand the systematic difference between the masses (luminosities) 
of open and globular clusters is less pronounced in the Large 
Cloud than it is in the Milky Way, i. e. the LMC open cluster
population contains a larger fraction of massive clusters 
than does that of Galactic open clusters. 

\section{THE EVOLUTION OF CLUSTER SYSTEMS}

\subsection{Unimodal and bimodal metallicity cluster systems.}

   It has been known for may years (van den Bergh 1975) that
the mean metallicities of globular cluster systems grows with 
increasing parent galaxy luminosity. This shows that the
nature of globular cluster systems is inextricably linked 
to the characteristics of their host galaxies. This leads
one to suspect that the properties of globular cluster
systems might also correlate with the Hubble types of their
parent galaxies. Unfortunately the number of globular cluster
systems for which detailed information on the metallicities 
of individual clusters has become available is still small.
The data are particularly incomplete for spiral galaxies.
However, Eerik \& Tenjes (2003) have recently published 
quite extensive information on the metallicity distributions
of globular cluster systems surrounding E7-E7 and S0 galaxies.
Their data appear to show (97.5\% confidence) that elliptical
galaxies are more likely to have bimodal cluster metallicity
distributions than is the case for S0 galaxies. However, this
effect appears to be due to the fact that that S0 galaxies 
are systematically fainter than ellipticals. On average 
ellipticals therefore have richer globular cluster systems than 
do lenticulars. The observed effect may be due to the fact that
it is difficult to establish bimodality of the cluster metallicity
distribution in a poor cluster systems. It would be very interesting 
to extend the study of the relationship between modality of cluster 
systems to spiral galaxies. A practical problem is, however, 
that spirals (1) have a lower specific globular clusters frequency 
than do ellipticals, and (2) that it is often difficult to 
distinguish unambiguously between stars and globular clusters in 
distant spirals.

\subsection{The flattening of globular clusters.}

   The majority of both open and globular clusters in the Galaxy
appear as little elongated almost spherical objects. A strikingly
different situation is observed to hold in the Large Magellanic 
Cloud (Frenk \& Fall 1982, van den Bergh \& Morbey 1984) in which 
both globular clusters and open clusters are found to be 
systematically more flattened than are those in the Milky Way System. 
Such highly flattened objects as NGC 121 in the SMC and NGC 1978 in the LMC
are very rare or absent in the Galaxy. The reason for this systematic
difference is presently not known. In this connection it is also of 
interest to note (van den Bergh 1996) that the most luminous globular 
in a cluster system (e. g. Omega Centauri in the Galaxy, Mayall II in
M31, NGC 1835 in the LMC and NGC 121 in the SMC) is often also the 
most highly flattened.

\section{DESIDERATA FOR FUTURE WORK}

   The topics covered in this review suggest that it might be
rewarding to study the following problems in more detail:

\begin{itemize}

\item Why is the age distribution of LMC globular clusters so 
      different from that of Galactic globulars? This difference
      is particularly puzzling because the dichotomy between the
      masses of open and globular star clusters appears to be much
      less pronounced in the Large Cloud than it is in the Galaxy.  
\item We need to understand why the evolutionary histories of
      different globular cluster systems differ. Why do some have
      bimodal metallicity distributions, whereas others do not.
\item Why are there systematic galaxy-to-galaxy differences between 
      the average flattenings of individual star clusters? Why do
      these mean flattenings depend on cluster luminosity, but not on 
      cluster age?
\item Some evidence suggests that the mass spectrum with which 
      stars form in clusters is universal, but other data indicate
      that some clusters may have been formed with ``top heavy''
      mass spectra. Resolving this problem would help us understand
      both the dynamical and chemical evolution of star clusters.  
\end{itemize}

   I also belatedly wish to thank Prof. ten Bruggencate for
giving me his personal copy of his book ``Sternhaufen'' just before 
his death. I am also indebted to Doug Johnstone for helpful
discussions.

\end{document}